\documentclass[floatfix,twocolumn,aps,prd,superscriptaddress,longbibliography]{revtex4-1}
\usepackage{graphicx}
\usepackage{dcolumn}
\usepackage{bm}
\usepackage{textcomp}
\usepackage[intlimits]{amsmath}
\usepackage{esint}
\usepackage{braket}
\usepackage{color}

\usepackage{soul} 

\soulregister\ref{7}
\soulregister\eqref{7}
\soulregister\cite{7}
\soulregister\onlinecite{7}

\usepackage{xspace}

\begin{document}
	
\title{Parallel random LiDAR with spatial multiplexing of a many-mode laser}

\author{Kyungduk Kim}
\author{Yaniv Eliezer}
\author{Olivier Spitz}
\author{Hui Cao}
\email[*]{hui.cao@yale.edu}
\affiliation{Department of Applied Physics, Yale University, New Haven, Connecticut 06520, USA}%

\begin{abstract}
    We propose and experimentally demonstrate parallel LiDAR using random intensity fluctuations from a highly multimode laser. We optimize a degenerate cavity to have many spatial modes lasing simultaneously with different frequencies. Their spatio-temporal beating creates ultrafast random intensity fluctuations, which are spatially demultiplexed to generate hundreds of uncorrelated time traces for parallel ranging. The bandwidth of each channel exceeds 10~GHz, leading to a ranging resolution better than 1 cm. Our parallel random LiDAR is robust to cross-channel interference, and will facilitate high-speed 3D sensing and imaging.
\end{abstract}

\maketitle 

\section{Introduction}

LiDAR (Light Detection and Ranging) is a key technology in autonomous driving, robotics, augmented and virtual reality. High-speed acquisition of distance information over a wide field of view is crucial for such applications~\cite{schwarz2010mapping}. The most common LiDAR scheme is based on time-of-flight measurement: the echo of an optical pulse tells the distance of a target from the time lag~\cite{amann2001laser}. Raster-scanning of a probe beam is commonly employed for three-dimensional (3D) ranging~\cite{kim2021nanophotonics, wang2020mems}. However, the mechanical scanning rate limits the acquisition speed. Parallel LiDAR based on multiple-input multiple-output (MIMO) scheme~\cite{fishler2004mimo} facilitates high-speed 3D mapping. To generate multiple probe beams in parallel, spectral and/or temporal multiplexing of broadband light sources~\cite{jiang2020time, riemensberger2020massively, lukashchuk2021chaotic, chen20223, zang2022ultrafast, chen2023breaking} are employed. With total bandwidth fixed, a larger number of spectral channels is accompanied by a narrower bandwidth of each channel, which means a worse resolution. Similarly, temporal multiplexing relies on interleaved transmission/reception, which limits the detection speed. Spatial multiplexing can avoid such problems while providing many more channels by using a 2D array of lasers~\cite{warren2018low, dummer2021role}. However, all spatial channels have similar temporal/spectral waveforms, which will cause channel interference and ranging ambiguity. 

The key issue for parallel ranging is to generate a large set of uncorrelated waveforms. For spatial multiplexing, one can use random time traces as fingerprints of individual channels. Previously, random modulation of continuous waves in time was employed for single-channel LiDAR~\cite{takeuchi1983random, nagasawa1990random, bashkansky2004rf, matthey2005pseudo, ai2011high, feng2021fpga, spollard2021mitigation, sambridge2021detection, chen2021investigation}, in order to avoid interference and jamming~\cite{guosui1999development, xu2001range, lin2004chaotic2, wang2017white}. Compared to pseudo-random sequences created by optical modulators, intensity fluctuations of chaotic lasers~\cite{myneni2001high, lin2004chaotic, ohtsubo2012semiconductor} and thermal emitters~\cite{zhu2012thermal}, as well as stochastic spontaneous emission events~\cite{tsai2020anti, hwang2020mutual}, can provide true random waveforms with higher modulation speed. In particular, a chaotic semiconductor laser with radio-frequency (RF) bandwidth $\sim$10 GHz provides centimeter resolution for long-distance ranging~\cite{lin2004chaotic, cheng20183d, chen20213d, ho2022high, tsay2023random}. However, it is technically challenging to make a large array of chaotic lasers with distinct dynamics for parallel LiDAR.

Here we propose and demonstrate parallel random LiDAR by using a single multimode laser for the simultaneous creation of many uncorrelated random time traces. Instead of using chaotic lasing dynamics, we resort to a different physical process---the spatiotemporal beating of many lasing modes in a single cavity---to generate ultrafast intensity fluctuations that vary spatially. In particular, a near-degenerate cavity can support lasing in many transverse and longitudinal modes. A slight detuning of the degenerate cavity lifts the frequency degeneracy of transverse modes, leading to a dense RF spectrum. A large number of transverse lasing modes offers several hundreds of probe beams with uncorrelated intensity fluctuations for parallel ranging. The resolution of an individual channel is determined by the lasing emission bandwidth, which is about 100 GHz for the solid-state laser used here. With sufficiently fast photodetection, the ideal resolution can be as fine as 1 mm, higher than that of chaotic semiconductor lasers. We further show that our parallel LiDAR scheme is robust to cross-interference between channels.

\section{Multimode lasing}

\begin{figure*}[t]
	\centering\includegraphics[width=0.6\linewidth]{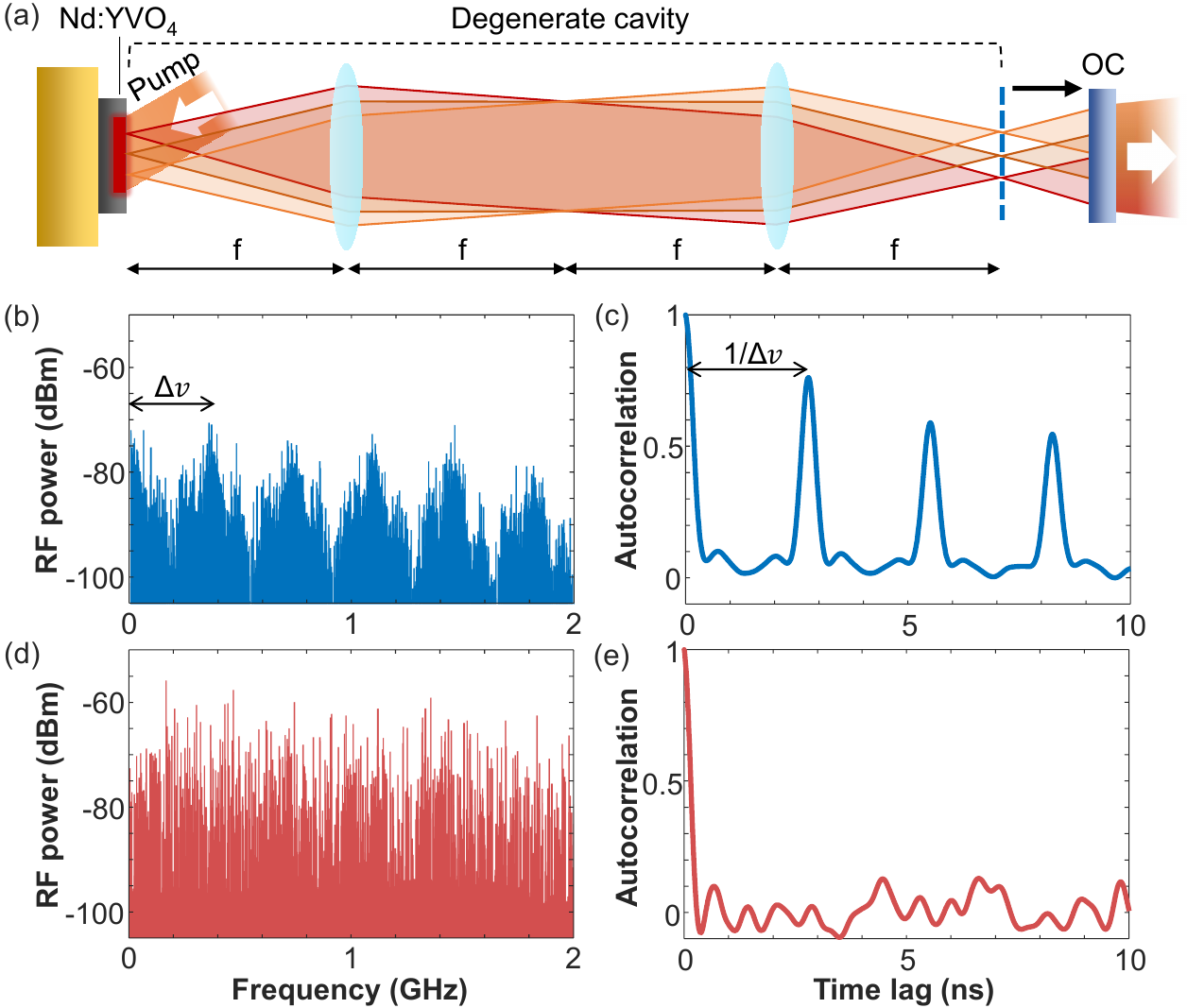}
	\caption{Detuning of a degenerate-cavity laser. (a) Schematic of a degenerate cavity laser. Between an end mirror and an output coupler (OC), two lenses are arranged in a self-imaging configuration. A thin Nd:YVO$_4$ disk is placed next to the end mirror and optically pumped to provide gain for lasing. The OC is moved slightly by about 1 cm (away from the dashed line) to lift the frequency degeneracy of transverse modes while keeping many transverse and longitudinal modes lasing. (b) RF spectrum of the nearly degenerate cavity, with peaks clustered around the integer multiples of longitudinal mode spacing (free spectral range) $\Delta\nu$ = 0.375 GHz. (c) Temporal correlation of emission intensity fluctuation before cavity detuning, featuring peaks at the integer multiples of $1/\Delta\nu$ = 2.7 ns. (d) RF spectrum after cavity detuning, with peaks more uniformly spread over frequency. (e) Temporal correlation of the emission intensity fluctuation after cavity detuning, showing long-range correlations are suppressed.}
	\label{fig1}
\end{figure*}

Instead of building a 2D array of separate lasers, we combine many lasers in a single cavity. This is realized with a degenerate cavity \cite{nixon2013efficient}, as shown schematically in Fig.~\ref{fig1}(a). It consists of two lenses in a $2f - 2f$ telescope arrangement between two flat mirrors \cite{liew2017intracavity}. The focal length $f$ of both lenses is 100 mm, which yields a cavity length $L = 4f = $ 400 mm. A thin disk of Nd:YVO$_4$ crystal, located at one end of the cavity, provides optical gain when optically pumped by a high-power laser diode at 808 nm (Coherent, FAP800-40W). The other end of the cavity is an output coupler with 99.6\% reflectivity at 1064 nm.

In principle, the degenerate cavity has the self-imaging property, where any point on one end mirror is imaged back onto itself after light propagates one round trip inside the cavity [Fig.~\ref{fig1}(a)]. This property, combined with optical pumping, allows simultaneous lasing in many transverse modes with degenerate frequency and quality factor \cite{arnaud1969degenerate, pole1965conjugate}. Each transverse mode creates a diffraction-limited area on the gain disk, and the total number of lasing modes is given by the ratio of the pump area over the diffraction area. The latter is determined by the numerical aperture of the cavity as well as optical misalignment. Small misalignments, aberrations, and thermal lensing induce a small degree of inherent detuning \cite{Arnaud69p2}, therefore in practice the frequency degeneracy is slightly lifted with limited spectral spread.

Continuous-wave lasing is achieved in the degenerate cavity with optical pumping at room temperature. The pump beam has a diameter of 2 mm on the gain disk. The optical pump power is 22 W, well above the lasing threshold of 4.1 W. The output emission power from the degenerate cavity is 1.1 W. Since the optical gain bandwidth ($> 100$ GHz) is much wider than the frequency spacing of longitudinal modes (free spectral range $\Delta\nu$ = 0.375 GHz), lasing occurs for hundreds of longitudinal modal groups, each containing many near-degenerate transverse modes \cite{chriki2018spatiotemporal}.

The beating of transverse and longitudinal lasing modes creates intensity variations in space and time. If the temporal fluctuation is random at each location and uncorrelated with other locations, these intensity traces can be used for parallel LiDAR via spatial multiplexing. We characterize the laser emission intensity fluctuation using an amplified InGaAs photodetector (Newport, 818-BB-30A, 1.5 GHz bandwidth) and a real-time high-speed oscilloscope (Keysight UXR0204A, 20 GHz bandwidth). We note that 1.5 GHz (20 GHz) is the minimum 3-dB bandwidth tabulated by the manufacturer, beyond which the signal is attenuated electrically (digitally). Fourier transform $\mathcal{F}$ of the time trace of emission intensity $I_t(t)$ gives the radio-frequency (RF) spectrum. Figure~\ref{fig1}(b) shows the experimental RF spectrum of the nearly degenerate cavity laser, obtained by taking the Fourier magnitude squared $|\mathcal{F}[I_t(t)]|^2$. It features periodic modulation, with dominant RF components clustered around the integer multiples of $\Delta \nu$. We also compute the temporal correlation of emission intensity, $C_t(\Delta t) = \langle I_t(t) \, I_t(t+ \Delta t) \rangle_t$, where $\langle ...\rangle_t$ denotes averaging over time $t$. In Fig.~\ref{fig1}(c), $C_t(\Delta t)$ exhibits a series of correlation peaks spaced by $\Delta t = 2.7$ ns, which corresponds to the round-trip time in the cavity. Such long-range correlation reflects periodic oscillation of emission intensity, which will significantly degrade the performance of our random LiDAR. 

\begin{figure*}[t]
	\centering\includegraphics[width=0.8\linewidth]{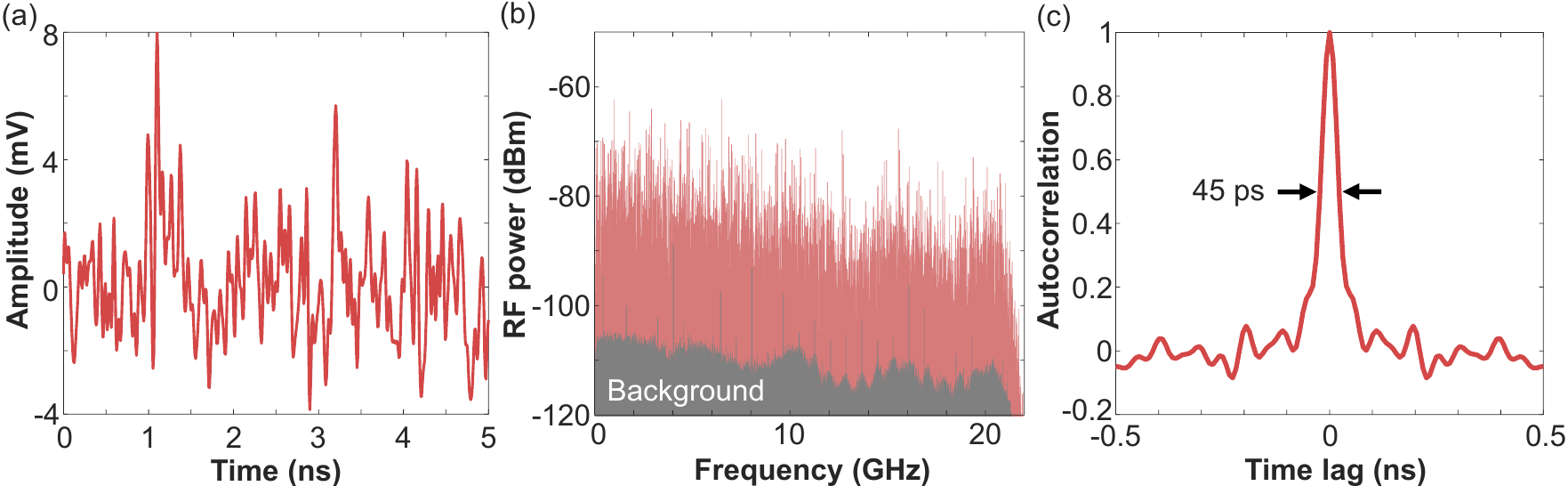}
	\caption{Axial resolution of random LiDAR. (a) Measured time trace of laser intensity fluctuation from the degenerate cavity with detuning. (b) RF spectrum of the emission intensity containing numerous frequency peaks over 20 GHz, as a result of beating among many different transverse and longitudinal lasing modes. (c) Temporal correlation of the emission intensity, revealing a short decorrelation time of 45 ps, which is limited by the temporal resolution of photodetection.}
	\label{fig2}
\end{figure*}

To suppress the long-range correlation in time, we need to generate a flatter RF spectrum, which is also a key feature in chaotic LiDAR ~\cite{lin2004chaotic}. This is achieved by further breaking the frequency degeneracy between different transverse lasing modes with cavity detuning \cite{mahler2021fast}. As shown in Fig.~\ref{fig1}(a), we slightly offset the axial location of the output coupler by about 1 cm. The transverse modes, represented by diffraction-limited spots at the original plane of the output coupler, will diffract and partially overlap at the new position. The spatial overlap leads to mode coupling, which lifts the frequency degeneracy. Temporal interference between these transverse modes will create additional beat notes that fill the gaps between integer multiples of $\Delta\nu$ in the RF spectrum [Fig.~\ref{fig1}(d)]. Consequently, long-range temporal correlations are significantly reduced, as shown in Fig.~\ref{fig1}(e). Hence, detuning the degenerate cavity geometry randomizes the temporal variation of laser intensity, which can be used for LiDAR.

\section{Axial resolution}

The axial resolution of random LiDAR is determined by the time scale of intensity fluctuation, or the bandwidth of the RF spectrum. To investigate how broad the RF spectrum is for the detuned cavity, we replace the aforementioned amplified photodetector with one having an order-of-magnitude higher bandwidth (Newport, 818-BB-36, 22 GHz bandwidth) combined with a low-noise amplifier (AT Microwave, AT-LNA-0018-3825H, 38-dB gain). The far-field emission out of the laser cavity is collected and sampled by a real-time high-speed oscilloscope (Keysight UXR0204A, 20 GHz bandwidth) at a sampling rate of 128 GS/s.

As shown in Fig~\ref{fig2}(a), the measured emission intensity exhibits rapid random fluctuation in time. The RF spectrum in Fig.~\ref{fig2}(b) features densely packed peaks caused by many-mode beating. Each peak has a narrow linewidth ($\sim$5 kHz measured from 1 ms-long trace, frequency resolution of 1 kHz). The numerous peaks are distributed more or less uniformly over a broad range in the RF domain. These peaks have irregular spacings and varying amplitudes, constituting a broad, dense RF spectrum. The sharp drop at 20 GHz is due to the limited bandwidth of the oscilloscope.

The wide RF spectrum corresponds to a short time scale of emission intensity fluctuation. Figure~\ref{fig2}(c) shows the temporal autocorrelation of intensity trace from the detuned degenerate cavity laser. The correlation width $\Delta t_c$, defined by full-width-at-half-maximum (FWHM) of the temporal correlation function $C_t(\Delta t)$, is 45 ps. It determines the axial resolution of the random LiDAR, namely, two targets separated axially by a distance $\Delta D = c\Delta t_c/2 = 7$ mm, can be resolved with $\Delta t_c = 45$ ps. 

The broad RF spectrum and fast temporal decorrelation of emission intensity from the detuned degenerate cavity laser will significantly improve the axial resolution of random LiDAR. The experimentally measured RF bandwidth and decorrelation time are still limited by the electrical bandwidth of our photodetection setup. The intrinsic RF spectrum width is dictated by the laser emission spectral width at optical frequency, which sets the largest beat notes (frequency difference) of lasing modes. We measure the optical spectrum of laser emission using a spectrometer with a wavelength resolution of 0.04 nm (Horiba TRIAX550 equipped with CCD3000, 1200 gr/mm). The wavelength width (FWHM) of the multimode laser emission is about 0.4 nm, corresponding to a frequency bandwidth of 100 GHz \cite{kim2021massively}. With sufficiently fast photodetection, the ultimate axial resolution of our LiDAR can be about 1 mm, which exceeds that of chaotic LiDAR.

\section{Spatial multiplexing}

The number of spatial channels available for parallel random LiDAR is determined by the number of transverse lasing modes. To ensure a large number of transverse modes lasing, the detuning of the degenerate cavity is kept small, so that the reduction of quality factors for high-order transverse modes is less significant. The beating of many transverse and longitudinal lasing modes creates complex intensity fluctuations in space and time. The number of distinct random waveforms generated at different spatial locations can be used as separate channels for parallel LiDAR.

\begin{figure}[t]
	\centering\includegraphics[width=\linewidth]{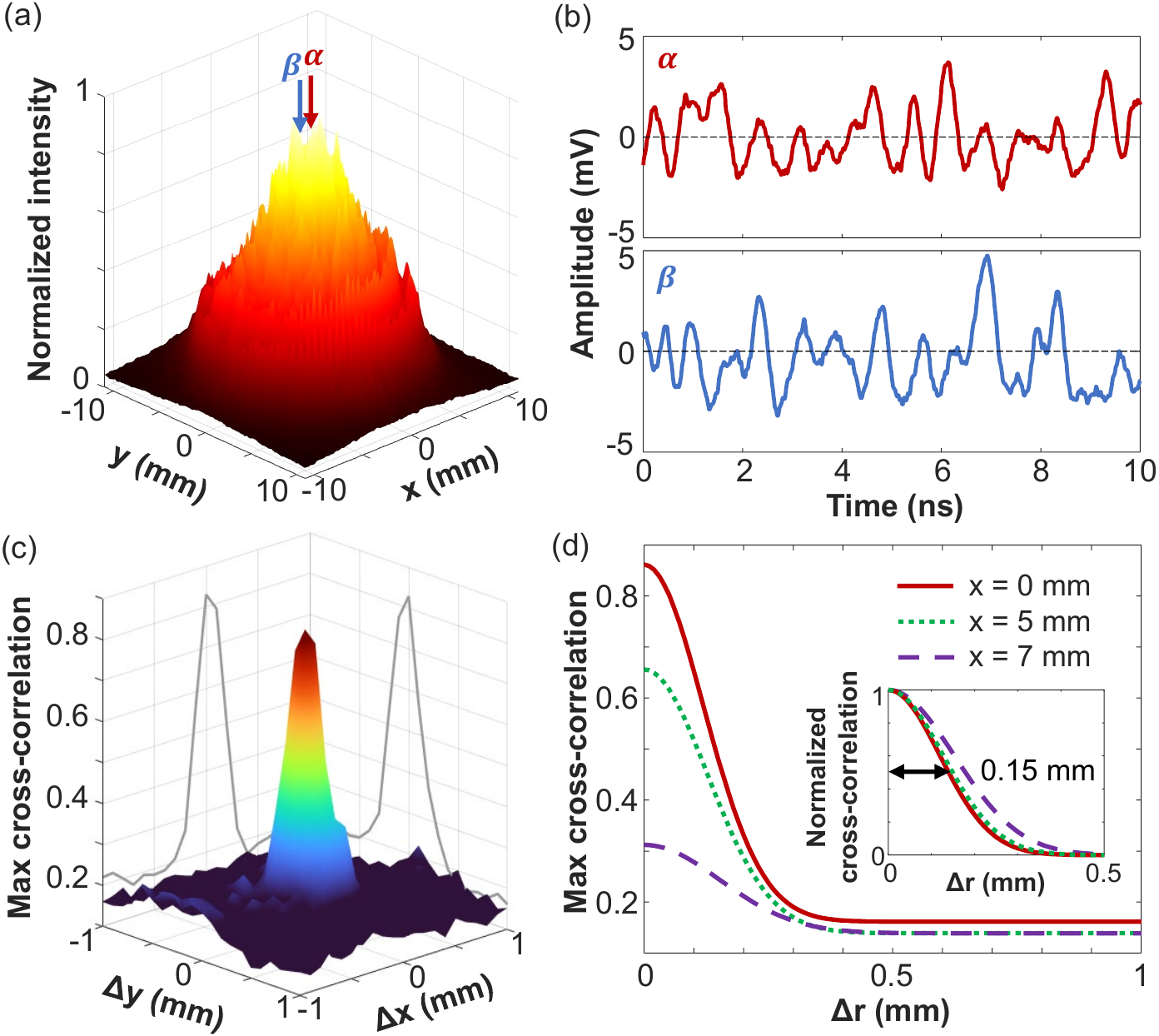}
	\caption{Spatial channels for parallel LiDAR. (a) Near-field intensity profile (time-integrated) of the laser emission from the detuned degenerate cavity. (b) Random fluctuations of emission intensity at two locations with transverse separation of 1 mm. (c) Maximal value of cross-correlation between intensity traces at different locations. Gray curves represent the correlations at $\Delta x$ = 0 and $\Delta y$ = 0. (d) Azimuthally averaged maximal cross-correlation versus radial distance $\Delta r = [(\Delta x)^2 + (\Delta y)^2]^{1/2}$, measured at different lateral positions $x$ on the beam while $y$ is fixed at 0 mm. Each curve represents a Gaussian fit of the distribution with a constant background. The peak of maximum cross-correlation decreases from the beam center ($x$ = 0 mm) to the edge ($x$ = 7 mm) due to a reduced signal-to-noise ratio. Inset: Normalized maximal cross-correlations, which have similar widths. The half-width-at-half-maximum increases slightly from 0.15 mm at $x$ = 0 mm to 0.18 mm at $x$ = 7 mm.}
	\label{fig3}
\end{figure}

To find the number of uncorrelated time traces generated by the detuned degenerate cavity laser,  we characterize the spatially-resolved temporal fluctuations of emission intensity. The laser beam profile at the output coupler is enlarged by a factor of 6.7 with a set of imaging optics. Figure~\ref{fig3}(a) shows the emission intensity profile measured by a CCD camera (Allied Vision, Mako G-131B). The time-integrated output pattern is a superposition of many transverse lasing modes. The effective beam diameter, defined by FWHM of azimuthally-averaged intensity, is 9.8 mm. The central region with strong emission within this effective beam diameter will be investigated for parallel LiDAR.

To measure intensity fluctuations at different spatial locations simultaneously, we divide the output laser beam using a 50/50 beamsplitter and place two amplified InGaAs photodetectors (Newport, 818-BB-30A, 1.5 GHz bandwidth) at each arm. The two arms have an identical length so that the two photodetectors simultaneously measure the intensity fluctuations at different locations of the output beam. By varying the transverse positions of the two photodetectors, we obtain two time traces with different separations across the beam. Figure~\ref{fig3}(b) shows temporal fluctuations of emission intensity at two locations with a transverse distance of 1 mm, denoted as $\alpha$ and $\beta$. The distinct time traces can be used as independent probes for parallel LiDAR.

To find how close two spatial channels can be, we characterize the spatial correlation of intensity fluctuations. For two transverse positions $(x,y)$ and $(x+ \Delta x, y+ \Delta y)$, the cross-correlation of  intensity fluctuations is given by $C_s(\Delta x, \Delta y; \Delta t) = \langle I(x, y, t) \, I(x+ \Delta x, y+ \Delta y, t+\Delta t) \rangle_{t}$. Experimentally, one photodetector is fixed in space, as the other one is scanned laterally across the output beam. For every offset $(\Delta x, \Delta y)$, the maximal magnitude of $|C_s(\Delta x, \Delta y; \Delta t)|$ among all time delay $\Delta t$ is plotted in Fig.~\ref{fig3}(c). The maximal cross-correlation has the largest value when the lateral positions of two photodetectors coincide ($\Delta x = \Delta y = 0$). With increasing offset in the lateral position, the time traces become different, and the maximal correlation drops rapidly. Eventually, at a large offset, the maximal cross-correlation levels off to a constant value. We attribute this residual correlation to the finite number of beat notes in the RF spectrum.

\begin{figure*}[t]
	\centering\includegraphics[width=0.75\linewidth]{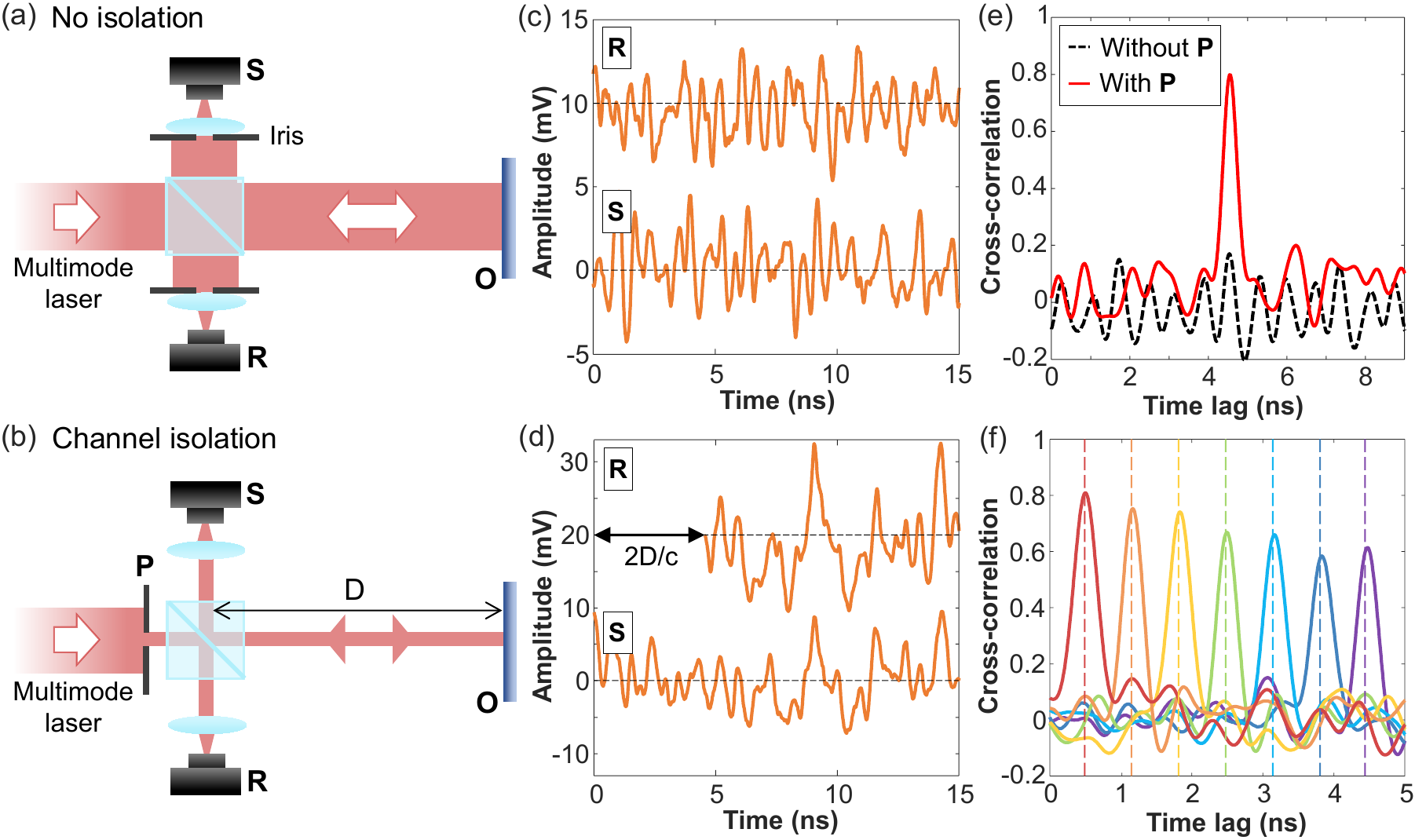}
	\caption{Isolated spatial channel for ranging. (a) Schematic of experimental setup without separating spatial channels in a multimode laser beam. The entire beam is split in two by a beamsplitter. The probe beam is reflected by a mirror \textbf{O} and focused by a lens to a photodetector \textbf{S}, while the reference beam is directed to another detector \textbf{R}. The irises in front of collecting lenses select the same spatial channel for \textbf{S} and \textbf{R}. (b) Schematic of a pinhole \textbf{P} in the laser beam to separate out a single channel. The setup is similar to (a), except the irises in front of the lenses are removed. (c) Measured signal $\textbf{S}$ has a completely different time trace from the reference $\textbf{R}$, without channel isolation in (a). The two traces are vertically offset for clarity. (d) Signal trace $\textbf{S}$ of an isolated channel in (b) is similar to the reference trace $\textbf{R}$ with a temporal offset of $2D/c$. (e) Cross-correlation between $\textbf{R}$ and $\textbf{S}$ in (d) features a dominant peak at the time lag $2D/c$ (solid red line). Without channel isolation, the cross-correlation (black dashed line) between $\textbf{R}$ and $\textbf{S}$ in (c) exhibits no peak at any time lag. Channel isolation is critical to our random LiDAR. (f) Cross-correlation between $\textbf{R}$ and $\textbf{S}$ with varying object distance $D$ (marked by vertical dashed line) for an isolated channel in (b). The time interval used to compute the cross-correlation is 200 ns.}
	\label{fig4}
\end{figure*}

The width of maximal cross-correlation sets the lateral width of a spatial channel of parallel LiDAR. Figure~\ref{fig3}(d) shows the azimuthally averaged maximal cross-correlation as a function of radial distance $\Delta r = [(\Delta x)^2 + (\Delta y)^2]^{1/2}$. Each curve is fit by a Gaussian function with a constant background. The three curves represent the cross-correlations measured at different locations across the beam: at the center ($x = 0$ mm, $y = 0$), at the effective beam radius ($x = 5$ mm, $y = 0$), and near the beam edge ($x = 7$ mm, $y = 0$). The peak value of correlation at $\Delta r = 0$ decreases from the beam center to the edge, as the emission intensity gets weaker and the signal-to-noise ratio (SNR) becomes lower (10 dB, 4.5 dB, 2.3 dB for $x$ = 0, 5, 7 mm, respectively). To find the correlation width, we normalize the correlation functions in the inset of Fig.~\ref{fig3}(d). The half-width-at-half-maximum (HWHM) increases from 0.15 mm at the beam center to 0.18 mm near the edge. The channel width $\Delta r_c$, defined as twice the HWHM, varies from 0.3 mm to 0.36 mm across the beam. To have spatial channels with uncorrelated intensity fluctuations, we set the channel spacing to 0.6 mm. Given the effective beam diameter of 9.8 mm [Fig.~\ref{fig3}(a)], the output beam of our single laser will provide approximately 300 independent spatial channels for parallel LiDAR.

\section{Isolating spatial channels}

One issue for long-distance ranging is the mixing of spatial channels outside the laser cavity. Upon axial propagation, individual channels diffract and overlap with other channels in space. Their interference will make temporal intensity fluctuations vary with propagation distance. Consider a spatial channel at transverse location $(x,y)$, its intensity trace at different longitudinal positions (along $z$-axis) is not related simply by a temporal offset, i.e., $I(x,y,z, t) \neq I(x,y,z+d, t+d/c)$, where $c$ is the speed of light. Instead, the temporal profile completely changes along $z$. 

Typically for ranging applications, the reference beam of one channel propagates a much shorter distance to a detector than the probe beam reflected from an object. Their intensity traces would be uncorrelated, as confirmed experimentally below. A multimode laser beam is split to reference and probe beams by a beamsplitter [see Fig.~\ref{fig4}(a)]. The reference beam is directed to a photodetector. An iris is placed in front of the lens to select a single spatial channel. The probe beam is reflected from a mirror and goes to a second detector. Another iris in front of the second lens selects the same spatial channel as the reference. The path length difference between the reference and probe beams is 1.35 m. Their intensity traces are completely different [Fig.~\ref{fig4}(c)], and exhibit no cross-correlation peak at any time lag [black dashed curve in Fig.~\ref{fig4}(e)]. This result indicates the temporal profile of intensity fluctuation in a single spatial channel becomes uncorrelated with free-space propagation, and thus cannot be used for ranging applications.

\begin{figure*}[t]
	\centering\includegraphics[width=0.8\linewidth]{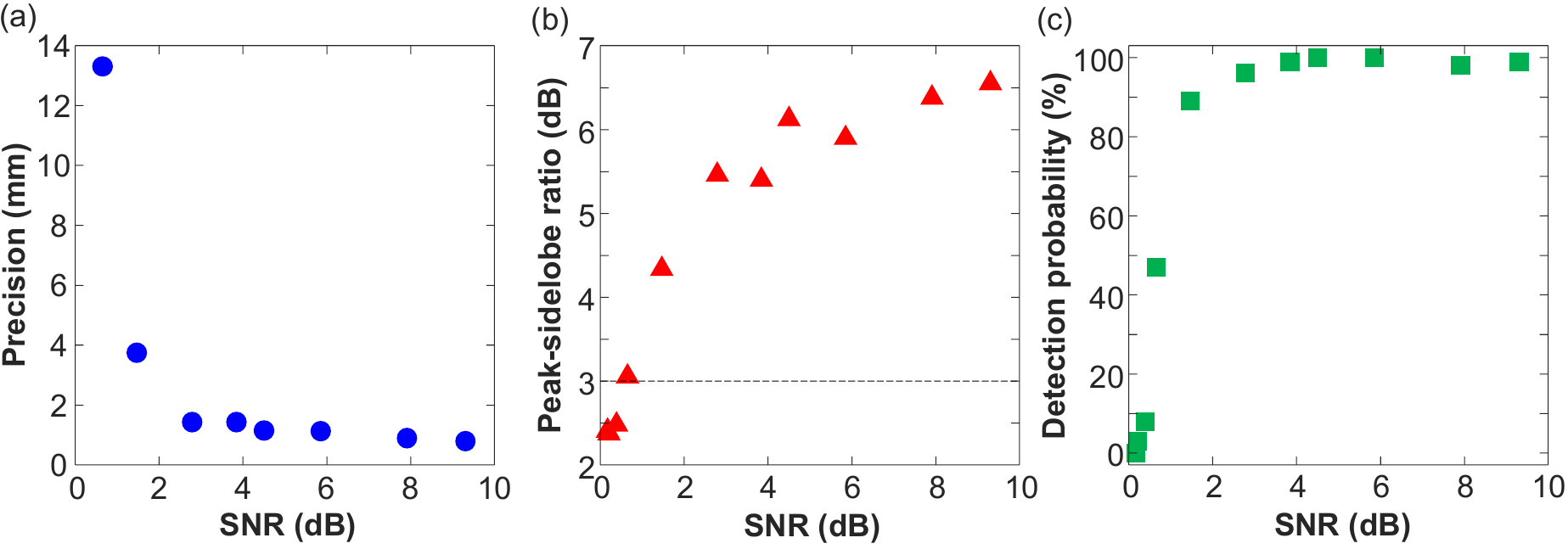}
	\caption{Random LiDAR performance with varying SNR. (a) The precision of the LiDAR as a function of SNR. It is better than 1.5 mm for SNR higher than 3 dB. (b) Peak-sidelobe ratio, indicating how strong the correlation peak is compared to the background. The black dashed line marks 3 dB, above which the detection is considered successful. (c) Detection probability of the measurement versus SNR. It remains nearly 100\% for SNR higher than 4 dB, and quickly degrades at lower SNR.}
	\label{fig5}
\end{figure*}

To resolve this issue, the spatial channels must be isolated to stop their beating outside the laser cavity. There are several methods of isolation. One is coupling the multimode laser emission into a fiber bundle by a lens array. The fields in individual single-mode fibers no longer couple, and the time traces of intensity remain invariant with propagation (except for a time lag). Another way is using a two-dimensional array of pinholes to isolate spatial channels. As a proof-of-concept demonstration, we filter a single spatial channel with a pinhole. Below we show its temporal profile remains invariant with axial propagation, and will use it for ranging.

Figure~\ref{fig4}(b) is a schematic of our experimental LiDAR setup with an isolated spatial channel. The output beam from the detuned degenerate cavity laser is attenuated to 0.3 W. Then, to select a single spatial channel, in the beam we place a pinhole with 
 0.5 mm diameter, which is slightly smaller than the channel spacing of 0.6 mm. The power of this separated single channel is 0.36 mW. The filtered beam passes through a 50/50 beamsplitter to form the reference and the probe. The reference beam directly goes to a photodetector \textbf{R}. As an object for LiDAR demonstration, we use a flat aluminium mirror with a reflectivity of 85\% at 1064 nm. The probe beam is reflected from the object and subsequently measured by a second photodetector \textbf{S}. To increase the collection efficiency, we use a plano-convex lens (focal length = 40 mm) to focus light onto the active area of the amplified photodetector (Newport, 818-BB-30A, 1.5 GHz bandwidth, 100 $\mu$m diameter). The probe beam power (after the 50/50 beamsplitter) is 0.18 mW, well below the safety regulation limit (class-1 lasers, 2 mW for emission at 1064 nm) for commercial LiDAR applications~\cite{dai2022requirements,ansi2014laser}.

Figure 4(d) shows the time traces measured by two photodetectors \textbf{S} and \textbf{R}. The signal trace \textbf{S} exhibits nearly identical intensity fluctuation as the reference \textbf{R}, except for an offset in time. The offset is given by $2D/c$, where $D$ is the distance from the beamsplitter to the object, as the distance from the beamsplitter to \textbf{S} is equal to that to \textbf{R}. The object distance $D$ can be retrieved from the cross-correlation between the signal and reference traces. Figure~\ref{fig4}(e) shows a dominant peak of the cross-correlation (solid red curve) at the time lag corresponding to $2D/c$. Experimentally we move the object so that $D$ varies from 7 cm to 66 cm, corresponding to an interval from 0.46~ns to 4.4~ns time delay. For all distances, a cross-correlation peak stands out of the fluctuating background [Fig.~\ref{fig4}(f)]. Its time lag gives the object distance $D$, which agrees well with the actual distance (vertical lines).

\section{Random LiDAR performance}

We characterize the axial resolution of the ranging measurement. The resolution is a measure to distinguish two adjacent peaks, and thus determined by the peak width of the cross-correlation. From Fig.~\ref{fig4}(f), the FWHM of the correlation peak $\Delta t_c= 0.35$ ns yields the axial resolution $c \, \Delta t_c/2$ = 5.3 cm. We note that the axial resolution is limited by the bandwidth of our photodetectors (1.5 GHz), and it can be improved with high-bandwidth photodetectors. 

We further measure the precision and detection capability of our LiDAR. The target is a high-reflectivity Aluminium mirror (reflectivity of 85\% at 1064 nm) placed at a distance $D$ = 30 cm. The sampling rate is 128 GS/s, and the time interval to compute cross-correlation is 200 ns. To characterize the LiDAR performance as a function of the SNR, we vary the signal strength by inserting neutral density filters into the beam path. To find the precision, we perform 100 consecutive ranging measurements and obtain statistics of the time lag for the maximal cross-correlation. Its standard deviation remains less than $\delta t = 10$ ps, which implies that the precision is $c \, \delta t/2$ = 1.5 mm, for SNR higher than 3 dB [Fig.~\ref{fig5}(a)]. However, the precision quickly becomes worse, once the SNR falls below 3 dB.

The detection capability of LiDAR is characterized by the peak-sidelobe ratio, which indicates how strong the peak is compared to the background of cross-correlation. It is defined as the ratio between the maximal peak height and three times the standard deviation of the background. We measure the peak-sidelobe ratio for 100 consecutive measurements, and the averaged results are plotted as a function of SNR in Fig.~\ref{fig5}(b). With SNR higher than 3 dB, the peak-sidelobe ratio remains larger than 5 dB, above the 3 dB limit for successful detection~\cite{chen20213d}.

Lastly, we investigate the detection probability [Fig.~\ref{fig5}(c)]. We repeat the measurement one hundred times and examine whether the target can be detected. The detection probability is defined by the fraction of successful detection over the total number of measurements. The detection probability remains nearly 100\% for the SNR higher than 4 dB. However, it rapidly drops at lower SNR.

\section{Parallel LiDAR}

\begin{figure}[b]
	\centering\includegraphics[width=\linewidth]{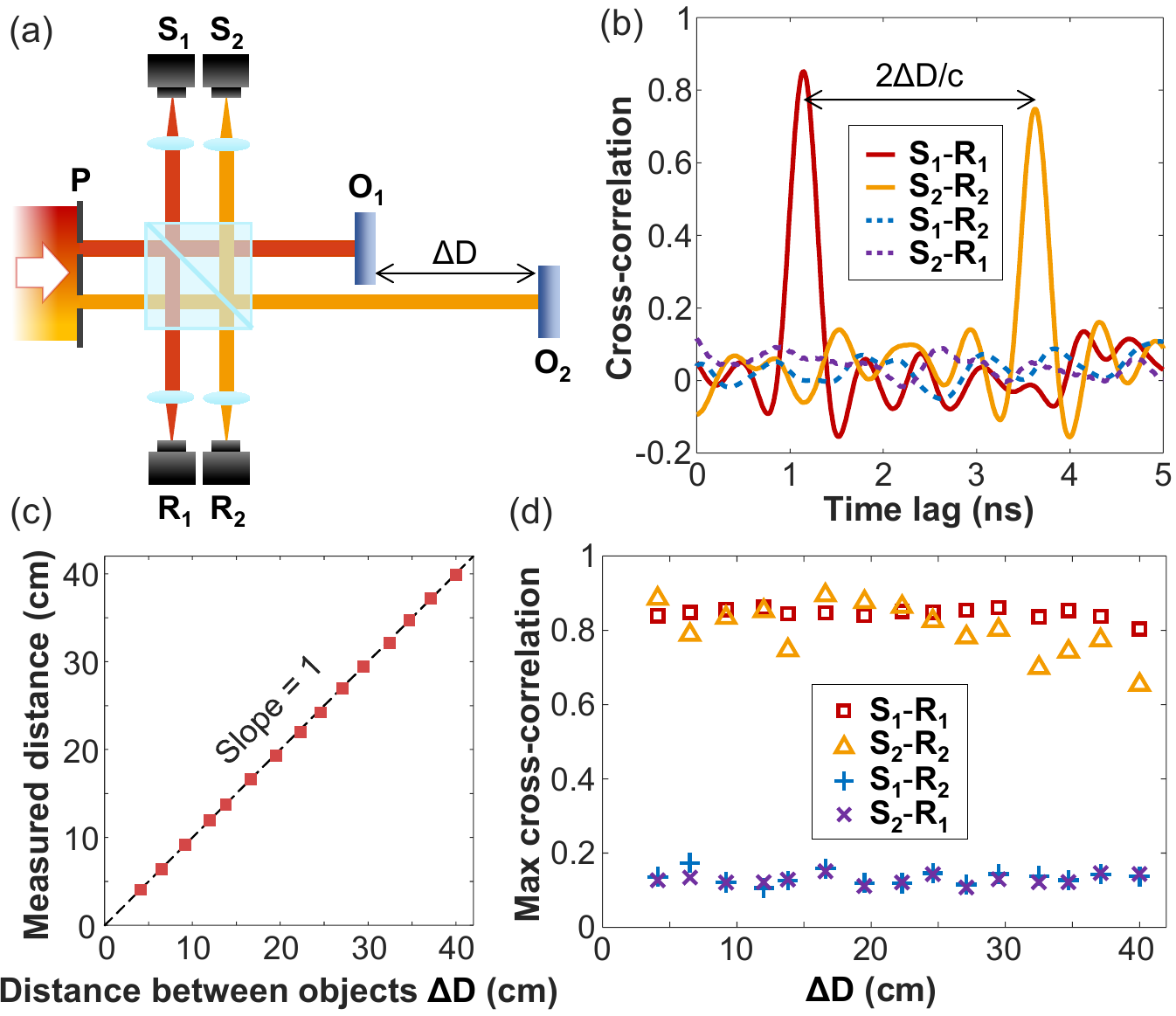}
	\caption{Parallel LiDAR with spatial multiplexing. (a) Schematic of ranging two objects $\mathbf{O_1}$ and $\mathbf{O_2}$ using two spatial channels. The laser emission passes through a dual-pinhole mask $\mathbf{P}$, which generates two beams with time traces invariant to axial propagation and uncorrelated to each other. A portion of beam serves as reference ($\mathbf{R_{1,2}}$), while another portion probes the objects $\mathbf{O_{1,2}}$ and return ($\mathbf{S_{1,2}}$). (b) The cross-correlation of time traces between every combination of $\mathbf{S_{1,2}}$ and $\mathbf{R_{1,2}}$. A peak is present only between the same channel. The distance between peaks $2\Delta D/c$ reveals the distance between two objects $\Delta D$ = 37 cm. (c) The measured distance with respect to the actual distance $\Delta D$, showing an excellent agreement. (d) For a wide range of $\Delta D$, the maximal cross-correlation between $\mathbf{S}$ and $\mathbf{R}$ remains high (low) for the same (different) channels.}
	\label{fig6}
\end{figure}

Next, we conduct simultaneous ranging of different targets using multiple spatial channels. As a proof-of-principle demonstration, we use two spatial channels. In order to generate two separate beams whose time traces are invariant with axial propagation, we use a double-pinhole mask $\mathbf{P}$ to isolate two spatial channels [Fig.~\ref{fig6}(a)]. Each pinhole diameter is 0.5 mm, and the edge-to-edge distance between the two pinholes is 2.0 mm. Since the double-pinhole mask is placed at the same plane where the spatial correlation function of laser emission in Fig.~\ref{fig3}(d) is measured, the spatial correlation width of 0.3 mm is much smaller than the separation of two pinholes, and the intensity fluctuations in two filtered channels are uncorrelated. Both channels are split to reference and probe beams. Two targets (aluminum mirrors with a reflectivity of 85\% at 1064 nm) are axially separated by 37 cm. Channel 1 (2) probes the first (second) target $\mathbf{O_1}$ ($\mathbf{O_2}$), and the difference in propagation distance between signal and reference beams is $D_1 = 16$ cm ($D_2 = 53$ cm). Both signal and reference of two spatial channels are measured simultaneously by four photodetectors (Newport, 818-BB-30A, bandwidth 1.5 GHz).

Figure~\ref{fig6}(b) shows the cross-correlations between the same or different channels. Between the signal and the reference of the same channel ($\mathbf{S_1}-\mathbf{R_1}$ and $\mathbf{S_2}-\mathbf{R_2}$), their cross-correlation clearly exhibits a peak. The peaks for the two channels are located at different time lags, as the distances to $\mathbf{O_1}$ and $\mathbf{O_2}$ are different. The spacing between two peaks is given by $2\Delta D/c$, where $\Delta D = |D_2 - D_1|$ is the axial distance between two objects. Figure~\ref{fig6}(b) reveals that $\Delta D$ is 37 cm, which matches the actual distance between $\mathbf{O_1}$ and $\mathbf{O_2}$. 

To show that our method can detect the objects with a wide range of distances, we vary $\Delta D$ by shifting one object $\mathbf{O_2}$ axially, while keeping the other object $\mathbf{O_1}$ fixed. Figure~\ref{fig6}(c) shows that the measured distance $\Delta D$ agrees well with the actual distance.

Lastly, we verify that two spatial channels are uncorrelated. Figure~\ref{fig6}(b) shows negligible cross-correlations between the signal and reference from different channels ($\mathbf{S_1}-\mathbf{R_2}$ and $\mathbf{S_2}-\mathbf{R_1}$). This is further confirmed at a varying distance between two objects $\Delta D$. As shown in Fig.~\ref{fig6}(d), over a wide range of $\Delta D$, the maximal cross-correlation for different channels remains low, while the maximal cross-correlation between signal and reference of the same channel is significantly higher. The absence of correlation between different spatial channels is a key to parallel random LiDAR.

\section{Robustness to interference}

Finally, we investigate the anti-interference capability of our parallel random LiDAR. Interference may occur when multiple beams from different spatial channels overlap in space due to a long free-space propagation distance, which can degrade the ranging performance. We start with the simplest case of overlapping fields $E_1$ and $E_2$ from two spatial channels,
	\begin{equation}
	\label{interfere}
    \begin{split}
	|E_1(t)+E_2(t)|^2 & = |E_1(t)|^2+|E_2(t)|^2+2\mathrm{Re}[E_1^*(t) \, E_2(t)] \\
    &= I_1(t)+I_2(t)+I_{1,2}(t),
    \end{split}
	\end{equation}
where $I_1(t)=|E_1(t)|^2$ and $I_2(t)=|E_2(t)|^2$ are field intensity of two channels, and $I_{1,2}(t)=2\mathrm{Re}[E_1^*(t) \, E_2(t)]$ represents their interference. For ranging with channel 1, the signal in Eq.~(\ref{interfere}) is cross-correlated with channel 1 reference [proportional to $I_1(t)$]. Here $I_2(t)$ and $I_{1,2}(t)$ cause additional fluctuations in channel 1 signal. Since the interference term $I_{1,2}(t)$ contains $E_1(t)$, it may lead to unwanted correlation with channel 1 reference $I_1(t)$.

We first experimentally investigate the interference of two spatial channels. Their respective probe beams, after being reflected from separate targets, become partially overlapped at the detector. In order to enhance their spatial overlap, we use two pinholes with 0.5 mm edge-to-edge distance to select two spatial channels, each pinhole diameter is still 0.5 mm. The experimental setup is modified from that of two-channel LiDAR in Fig.~\ref{fig6}(a). The axial distance from the beamsplitter to object $\mathbf{O_1}$ is 17 cm, and to object $\mathbf{O_2}$ is 26 cm. Thus the two objects are separated axially by 9 cm. The two signal beams hit $\mathbf{O_1}$ and $\mathbf{O_2}$ separately. The reflected beams $\mathbf{S_1}$ and $\mathbf{S_2}$ diffract and partially overlap at a photodetector. The measured signal $\mathbf{S_{1+2}}$ is a mixture of $\mathbf{S_1}$ and $\mathbf{S_2}$. 

\begin{figure}[b]
	\centering\includegraphics[width=\linewidth]{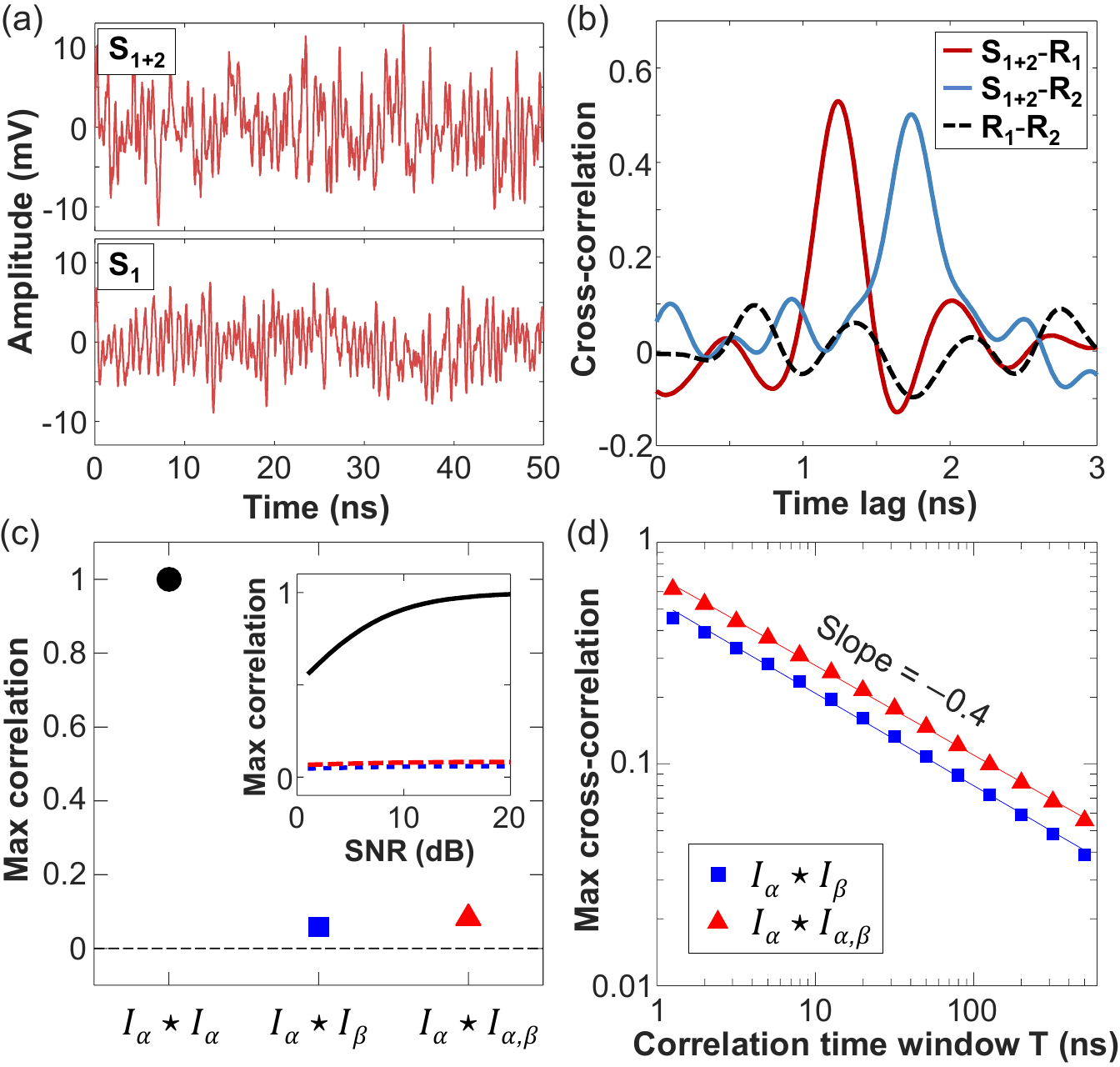}
	\caption{Interference of two spatial channels. (a) Experimentally measured time trace of mixed echoes $\mathbf{S_{1+2}}$ of two objects $\mathbf{O_1}$ and $\mathbf{O_2}$, probed by separate spatial channels 1 and 2 (upper panel). The magnitude of intensity fluctuation is larger than the echo $\mathbf{S_{1}}$ of channel 1 from $\mathbf{O_1}$ without $\mathbf{O_2}$ (lower panel). (b) Cross-correlation between the mixed signal $\mathbf{S_{1+2}}$ and the reference of individual channel $\mathbf{R_1}$, $\mathbf{R_2}$. Despite notable channel interference, strong correlation peaks reveal the axial distances of $\mathbf{O_1}$ and $\mathbf{O_2}$. The time interval for computing the correlation is $T = 200$ ns. (c) Maximal correlations of numerically simulated time traces, showing auto-correlation of intensity fluctuation in a single channel $I_{\alpha} \star I_{\alpha} =1$, and cross-correlation between two spatial channels  $I_{\alpha} \star I_{\beta} \simeq 0$. The correlation between a single-channel intensity $I_{\alpha}$ and its interference with another channel $I_{\alpha,\beta}$ is close to 0. All correlations are computed within 200 ns interval and averaged over 15 spatial channels. Inset: Simulated effect of detection noise on correlation as a function of SNR. $I_{\alpha} \star I_{\alpha}$ (solid black) decreases with SNR, but $I_{\alpha} \star I_{\beta}$ (dotted blue) and $I_{\alpha} \star I_{\alpha,\beta}$ (dashed red) do not change significantly. (d) Simulated maximal correlations $I_{\alpha}\star I_{\beta} \simeq 0$ and $I_{\alpha}\star I_{\alpha,\beta}$, decaying with increasing accumulation time $T$ of correlation in the absence of detection noise. The straight lines represent a linear fit in the log-log scale.}
	\label{fig7}
\end{figure}

As shown in Fig.~\ref{fig7}(a), the time trace of $\mathbf{S_{1+2}}$ features a larger fluctuation amplitude than $\mathbf{S_1}$, which is the echo of $\mathbf{O_1}$ with $\mathbf{O_2}$ removed. 
The SNR of two mixed channels $\mathbf{S_{1+2}}$ is 12 dB, while that for one channel $\mathbf{S_1}$ is 9 dB, which are both far below the 33 dB SNR at the saturation limit of our photodetector. The difference between $\mathbf{S_{1+2}}$ and $\mathbf{S_1}$ is 3 dB, implying that the fluctuation power for $\mathbf{S_{1+2}}$ is twice of that for $\mathbf{S_1}$. It indicates that the echoes $\mathbf{S_1}$ and $\mathbf{S_2}$ from two objects $\mathbf{O_1}$ and $\mathbf{O_2}$ are mixed with nearly identical amplitude of fluctuation.

Figure~\ref{fig7}(b) shows cross-correlations between measured time traces. First of all, the two references $\mathbf{R_1}$ and $\mathbf{R_2}$ have negligible correlation, confirming the two spatial channels are uncorrelated. The mixed signal $\mathbf{S_{1+2}}$ has correlation peaks with both references $\mathbf{R_1}$ and $\mathbf{R_2}$, but at different time lags, which correspond to the axial distances of objects $\mathbf{O_1}$ and $\mathbf{O_2}$. Hence, both objects are clearly distinguished despite the spatial overlap of their echoes.

To understand the experimental results, we numerically simulate cross-channel interference. A multimode laser beam is constructed by linear superposition of many transverse and longitudinal modes (see Appendix for details). The transverse mode frequencies are randomly distributed over the free spectral range. We choose 15 spatial channels at different positions $\alpha=1,\cdots,15$, and compute the time trace of the optical field in each channel $E_\alpha(t)$. From it, we calculate the intensity trace $I_\alpha(t) = |E_\alpha(t)|^2$ as well as its interference with another channel $I_{\alpha,\beta}(t) =  2\mathrm{Re}[E^*_{\alpha}(t) \, E_{\beta}(t)]$ [see Eq.~(\ref{interfere})]. The mean intensity $\langle I_\alpha(t) \rangle_t$ is set to be identical for all channels. For every interference term $I_{\alpha,\beta}(t)$, two channels $\alpha$ and $\beta$ are mixed with equal amplitude.

We calculate the cross-correlation for every pair of spatial channels, and average the maximal cross-correlation over all possible pairs of 15 channels. The accumulation time for correlation is set to 200 ns, to be consistent with the experiment. Figure~\ref{fig7}(c) shows that cross-correlation between different channels $I_{\alpha} \star I_{\beta}$ is close to zero, while autocorrelation of every channel $I_{\alpha} \star I_{\alpha}$ is unity. Moreover, the correlation between the intensity trace of a single channel and its interference with another channel, $I_{\alpha} \star I_{\alpha,\beta}$, also vanishes. Hence, the interference between spatial channels does not create an additional peak in the correlation between signal ($I_{\alpha} + I_{\beta} + I_{\alpha,\beta}$) and reference ($I_{\alpha}$) of the individual channel.

The channel interference effects diminish as a result of averaging correlation over a sufficient time interval $T$. Even when the interference term $I_{\alpha,\beta}(t)$ has a comparable magnitude to the signal $I_{\alpha}(t)$, its correlation with the reference $I_{\alpha}(t)$ decreases monotonically with increasing accumulation time $T$. In Fig.~\ref{fig7}(d), maximal correlations of both $I_{\alpha} \star I_{\beta}$ and $I_{\alpha} \star I_{\alpha,\beta}$ decay with a similar slope in the double logarithmic scale. 

The numerical results in Figs.~\ref{fig7}(c,d) are obtained without noise, yet in real measurements the detection noise plays a significant role. Based on the measured noise of our photodetector, we simulate the detection noise with uncorrelated Gaussian intensity distribution. The inset of Fig.~\ref{fig7}(c) shows that the noise has a notable impact on the autocorrelation  ($I_{\alpha} \star I_{\alpha}$), which decreases with reducing SNR. The cross-correlation ($I_{\alpha} \star I_{\beta}$) and the interference term ($I_{\alpha} \star I_{\alpha,\beta}$) are less affected by the noise, as they are already uncorrelated before adding the noise. The entire curves of maximal cross-correlation in Fig.~\ref{fig7}(d) shift downward in the presence of noise but its slopes are unchanged (not shown).

When the number of overlapping channels is more than two, the mixed echo with equal contribution from $M$ channels is expressed as: 
\begin{equation}
\label{interfereM}
\begin{split}
\bigg |\sum_{\beta=1}^M E_\beta(t) \bigg |^2 &  = I_\alpha(t) +  \sum_{\beta \neq \alpha} I_\beta(t) + \sum_\gamma \sum_{\beta \neq \gamma} E_\gamma^*(t) \, E_\beta(t) \\
& = I_\alpha(t) + I_{\rm inc}(t) + I_{\rm coh}(t) \, .
\end{split}
\end{equation}
$I_{\rm inc}(t)$ denotes an intensity sum of all channels except $\alpha$, and $ I_{\rm coh}(t)$ includes field interference terms among all channels. Correlating the mixed signal with channel $\alpha$ reference produces three terms. The correlation peak of the first term gives the distance of an object probed by channel $\alpha$, and the second and third terms add fluctuating backgrounds to cross-correlation. Our numerical results reveal that the maximum of the second term $I_\alpha \star I_{\rm inc}$ scales as $\sqrt{M}$ for $M \gg 1$, and the maximum of the third term $I_\alpha \star I_{\rm coh}$ scales as approximately $M$. Both drop with accumulation time $T$, making background correlations negligible for sufficiently long $T$. Hence, our parallel LiDAR is robust to cross-channel interference.

\section{Discussion and conclusion}

In summary, we experimentally demonstrate parallel random LiDAR by utilizing spatiotemporal intensity fluctuations from a highly multimode laser. By detuning a degenerate cavity, we lift the frequency degeneracy of transverse modes without significant reduction of quality factors, so that lasing occurs in several hundreds of transverse modes simultaneously. Thanks to their frequency diversity, the spatio-temporal interference of all lasing modes creates numerous beat frequencies densely packed in the RF domain. The complex intensity fluctuations in space and time produce several hundreds of probe beams for parallel ranging. Our method does not need any optical modulator or external feedback. Instead, a single free-running laser can generate a large set of random waveforms in parallel. For the proof-of-concept demonstration, we have used an existing degenerate cavity laser of length 40 cm. For applications such as autonomous vehicles, robots, drones, compact size is essential, and our laser cavity length may be halved by placing a mirror at the cavity center, and further reduced to a few centimeters by using a lens of shorter focal length, while keeping the same number of transverse lasing modes. The output power of our degenerate cavity laser can reach 10 W with sufficient cooling \cite{liew2017intracavity}. To further increase the power, a single multimode amplifier may be used to amplify the multimode laser emission.   

To keep the temporal profiles of intensity fluctuations in individual spatial channels invariant with free-space propagation, we stop the spatio-temporal mode beating outside the laser cavity by isolating spatial channels with pinholes. The number of parallel spatial channels with uncorrelated intensity fluctuations is limited by the number of transverse lasing modes~\cite{kim2021massively}. We expect 300 spatial channels would be available from a single laser. In this proof-of-principle demonstration, we use only two spatial channels, but it will be straightforward to further increase the number of spatial channels by using a 2D array of pinholes or coupling the emission into a fiber bundle. 
$M$ collimated beams may be steered simultaneously by a single mechanical element for block scanning. Compared to the temporal or spectral multiplexing scheme which uses a single photodetector in the receiver, our spatial multiplexing scheme requires an array of $M$ photodetectors to measure the time traces of individual spatial channels simultaneously. Alternatively, a single photodetector may record the signals from multiple spatial channels, thanks to uncorrelated intensity fluctuations in different channels [Figs.~\ref{fig7}(a,b)].
	
The broad and dense RF bandwidth of highly multimode laser emission enables a superior axial resolution of LiDAR. In our two-channel demonstration, the photodetectors have a bandwidth of 1.5 GHz, which limits the axial resolution to 5.3 cm. By using the 22-GHz bandwidth photodetector, the axial resolution can be further improved to 0.7 cm. The axial resolution is ultimately limited by the optical spectral bandwidth of the laser emission. Given the measured multimode emission bandwidth of 100 GHz, the axial resolution of our LiDAR can be as high as 1 mm. In practical situations of limited detection bandwidth, a narrowband spectral filter may be inserted into the degenerate cavity to reduce the laser emission linewidth in order to increase the detection efficiency. 

We also demonstrate the robustness of our parallel ranging scheme to cross-channel interference. Although the mixing of signals from different spatial channels significantly changes the temporal profile of intensity fluctuations, cross-correlation with the reference of the individual channel can distinguish different targets. This is because the cross-channel interference terms have a negligible correlation with the intensity fluctuation of individual channels. Such anti-interference characteristics will be particularly useful in the presence of many-channel interference, making our method suitable for massively parallel LiDAR.

In addition to LiDAR, our scheme is applicable to parallel RADAR. Direct measurement of laser emission from a detuned degenerate cavity by a 2D array of photodiodes will generate distinct random waveforms to feed an array of antennas for spatially demultiplexed RADAR. Our scheme has advantages over other demultiplexing schemes of chaotic signals in radio-frequency~\cite{cheng2015generation}, optical frequency~\cite{yao2015distributed, lukashchuk2021chaotic,lukashchuk2022chaotic}, polarization~\cite{zhong2017real, zhong2021precise}, and time domain~\cite{chen20223, feng2022pulsed}. Time-division demultiplexing inevitably requires a longer acquisition time for more channels, while for spatial demultiplexing the number of channels is independent of acquisition time. Compared to spectral demultiplexing, spatial demultiplexing is able to provide many channels without sacrificing the bandwidth and resolution of the individual channel. Moreover, our source has an intrinsic bandwidth of 100 GHz and can thus generate simultaneously several independent channels in various RF bands of interest for RADAR, such as the $\mathrm{K_u}$ band (12-18 GHz), the $\mathrm{K_a}$ band (27-40 GHz) and the $\mathrm{W}$ band (75-110 GHz), where the last is not accessible with temporal chaos from semiconductor lasers under optical feedback~\cite{lukin2022generation}. Without those trade-offs, our parallel ranging scheme based on a high-power many-mode laser will facilitate high-speed 3D sensing and imaging.

\setcounter{section}{0}
\renewcommand\thesection{\Alph{section}}
\section*{Appendix: Numerical modeling}

\subsection*{Multimode lasing}

To simulate the laser beam in a detuned degenerate cavity, we calculate a linear superposition of many spatial modes with randomly distributed frequencies. Considering the cylindrical symmetry of the laser cavity, we choose Laguerre-Gaussian modes as the modal basis. The spatial profile of a Laguerre-Gaussian mode in a polar coordinate $\mathbf{r} = (\rho,\theta)$ is given by~\cite{siegman1986lasers},
\begin{equation}
E_{pl}(\mathbf{r}) \propto \left( \frac{\sqrt{2} \rho}{w_0} \right)^{|l|} L_p^{|l|}\left( \frac{2\rho^2}{w_0^2} \right) e^{-\rho^2/w_0^2} e^{il\theta},
\end{equation}
where $p$ and $l$ are the radial and azimuthal mode numbers, $L_p^{|l|}(\cdot)$ is the generalized Laguerre polynomial, and $w_0$ is the beam radius for the fundamental transverse mode. Every modal profile is normalized by $\int |E_{pl}(\mathbf{r})|^2 d\mathbf{r} = 1$. We sort the modes $E_{pl}(\mathbf{r})$ from low-order to high-order transverse modes in ascending order of $2p+|l|$, and label them by $E_{m}(\mathbf{r})$ with the transverse mode index $m$. For instance, $m=0,1,2,3,4,5$ correspond to $(p,l)$ of (0,0), (0,1), (0,-1), (1,0), (0,2), (0,-2), respectively.

A linear superposition of many transverse and longitudinal modes is expressed as:
\begin{equation}
E(\mathbf{r},t)=\sum_{\substack{0\leq m<M \\ q_{\mathrm{min}}\leq q<q_{\mathrm{max}}}} E_{m}(\mathbf{r}) e^{-i(2\pi\nu_{m,q}t+\phi_{m,q})},
\end{equation}
where $q$ is the longitudinal mode index, $\nu_{m,q}$ is the resonant mode frequency, and $\phi_{m,q}$ is the random phase between 0 and $2\pi$. We sum over $M=300$ Laguerre-Gaussian transverse modes, and $q_{\mathrm{max}}-q_{\mathrm{min}}=40$ longitudinal modal groups. The lowest frequency $\nu_{0,q_{\mathrm{min}}}=c/{\lambda_0}$ is set by the emission wavelength $\lambda_0 = $ 1064 nm. The free spectral range $\Delta\nu_q = \nu_{0,q+1}-\nu_{0,q}$ is given by 375 MHz. 

To simulate the detuning of a degenerate cavity, we randomize the distribution of transverse mode frequencies. The spacing between adjacent transverse modes $\Delta\nu_{m,q} = \nu_{m+1,q}-\nu_{m,q}$ follows a uniform probability distribution between 0 and $2\langle \Delta \nu_{m,q} \rangle$. The mean transverse mode spacing $\langle \Delta \nu_{m,q} \rangle$ is set to $\Delta\nu_q/M$, which makes the transverse modes almost uniformly distributed within one free spectral range.

After calculating the near-field profile $E(\mathbf{r},t)$, we extract 15 time traces at a normalized distance $r/w_0=$ 2.9 from the cavity axis. These locations have the largest number of overlapping transverse modes, which leads to minimal long-range temporal correlations of intensity fluctuations. The time step of sampling is set to 20 ps. For Figs.~\ref{fig7}(c-d), we average over 10 random distributions of mode frequencies.

\subsection*{Cross-correlation}

We first normalize the field $E_{\alpha}(t)$ at every channel $\alpha$ such that the variance of the intensity fluctuation $\delta I_{\alpha}(t) = I_{\alpha}(t) - \langle I_{\alpha}(t) \rangle_t$ becomes unity, i.e. $\langle [\delta I_{\alpha}(t)]^2 \rangle_t = 1$, where $I_{\alpha}(t) = |E_{\alpha}(t)|^2$ is the intensity time trace. The maximal cross-correlation between two intensity time traces $I_{\alpha}(t)$ and $I_{\beta}(t)$ are then calculated as,
\begin{equation}
I_{\alpha} \star I_{\beta} = \max_{\Delta t} \langle \delta I_{\alpha}(t) \delta I_{\beta}(t+\Delta t) \rangle_t,
\label{xcorr}
\end{equation}
where $\langle$...$\rangle_t$ is the average within the time window of length $T$. The cross-correlation with any other intensity trace $I_x(t)$ [$I_{\alpha,\beta}$, $I_{\textrm{inc}}$, and $I_{\textrm{coh}}$ of the main text] is also calculated by Eq.~(\ref{xcorr}) but with $\delta I_{\beta}(t)$ replaced by $\delta I_x(t) = I_x(t) - \langle I_x(t) \rangle_t$. 

The intensity time trace of noise is simulated by the independent Gaussian distribution, with the mean $\mu =$ 0 and the standard deviation $\sigma = 10^{-(s/20)}$, where $s$ is the SNR in decibels. It is added to the intensity trace of signal $I_x(t)$, and the variance of the intensity fluctuation is normalized before plugging into Eq.~\ref{xcorr}.

\section*{Acknowledgement}
We acknowledge funding from US Naval Office of Research under Grant No. N00014-221-1-2026.
\bigskip\bigskip\bigskip\bigskip


\end{document}